\documentclass[10pt,aps,superscriptaddress,showpacs,pra,nofootinbib,twocolumn]{revtex4-1}
\usepackage{bm} 
\usepackage{graphicx} 
\usepackage{amsmath}
\usepackage{amsthm}
\usepackage{amssymb} 
\usepackage{epstopdf} 
\usepackage{amsfonts}
\usepackage{epsfig}
\usepackage{comment}
\usepackage{color} 
\usepackage{bm} 
\usepackage{subfigure}
\usepackage[english]{babel}

\newcommand {\be}{\begin{equation}}
\newcommand {\ee}{\end{equation}}

\newcommand\D{{\text{c}}}

\newcommand{\ba}{\begin{eqnarray}}
\newcommand{\ea}{\end{eqnarray}}
\newcommand\tr{{\mbox{Tr\,}}}
\newcommand{\ignore}[1]{}

 \usepackage[bb=boondox]{mathalfa}

\usepackage[margin=1in,nofoot]{geometry}

\newcommand{\e}{{{e}}}
\newcommand{\x}{{E}}
\newcommand{\rmd}{{\text d}}

\newcommand{\beq}{\begin{equation}}
\newcommand{\eeq}{\end{equation}}
\newcommand{\beqnn}{\begin{equation*}}
\newcommand{\eeqnn}{\end{equation*}}
\newcommand{\bea}{\begin{eqnarray}}
\newcommand{\eea}{\end{eqnarray}}
\newcommand{\beann}{\begin{eqnarray*}}
\newcommand{\eeann}{\end{eqnarray*}}
\newcommand{\bes} {\begin{subequations}}
\newcommand{\ees} {\end{subequations}}

\begin{document}

\title{Off-Diagonal Series Expansion for Quantum Partition Functions}

\author{Itay Hen}
\email{itayhen@isi.edu}
\affiliation{Information Sciences Institute, University of Southern California, Marina del Rey, California 90292, USA}
\affiliation{Department of Physics and Astronomy and Center for Quantum Information Science \& Technology, University of Southern California, Los Angeles, California 90089, USA}

\date{\today}

\begin{abstract}
We derive an integral-free thermodynamic perturbation series expansion for quantum partition functions which enables an analytical term-by-term calculation of the series. The expansion is carried out around the partition function of the classical component of the Hamiltonian with the expansion parameter being the strength of the off-diagonal, or quantum, portion. To demonstrate the usefulness of the technique we analytically compute to third order the partition functions of the 1D Ising model with longitudinal and transverse fields, and the quantum 1D Heisenberg model. 
\end{abstract}

\maketitle

\section{Introduction}
In statistical mechanics, all the thermodynamic functions can be expressed in terms of the system's partition function~\cite{AIC:AIC690190547,LL:SP1,LL:SP2}. 
However only a handful of many-body models 
admit analytical closed-form expressions for such a fundamental quantity. Examples for exactly solvable classical many-body systems are the Ising model in one and two dimensions (the latter in the absence of external fields)~\cite{Baxter:book}. Quantum systems of interacting particles that admit closed-form expressions for their partition functions are even rarer~\cite{Lieb:61,Pfeuty:70}. 

In the absence of closed-form expressions, exact-numerical methods such as quantum Monte Carlo are often used to statistically sample the partition function. 
Many models of physical interest are however difficult to evaluate even approximately in this way, especially in the thermodynamic limit~\cite{Loh-PRB-90,Troyer2005}. In this case, one normally resorts to perturbative methods and other approximation schemes, which have their specific ranges of applicability as well~\cite{seriesExpBook}. Among these are low-temperature series expansions~\cite{lowTempExpansion}, high-temperature expansions~\cite{BETTS1969150,FERNANDEZ19921283,0022-3719-4-15-023} and other types of series~\cite{Zhou2011,Zhou2012}.   

Here, we propose an integral-free thermodynamic perturbation scheme for the exact term-by-term calculation of the partition function of quantum many-body systems that is based on a series expansion in the `off-diagonal' coupling strength of the system in question. As we show, this technique allows for a relatively simple analytical evaluation of the quantum partition function in growing orders of quantum strength. 

The present approach is founded on a high-temperature Taylor series expansion of the partition function followed by the regrouping, or contraction, of terms of the same off-diagonal order, utilizing the concept of `divided differences'~\cite{dd:67,deboor:05,ODE}---which
in turn leads to a formulation of the quantum partition function as a series in the strength of its off-diagonal component and at the same time obviates the need for performing (sometimes cumbersome) multidimensional integrals in imaginary time as in standard thermodynamic perturbation theory~\cite{seriesExpBook}. 
We further argue that the suggested expansion naturally admits a simple diagrammatic depiction. We illustrate the applicability of the technique by calculating coefficients up to the third order of the quantum 1D Ising model and the quantum 1D Heisenberg model in the zero-magnetization sector. We also discuss additional potential uses of the technique as well as its relation to Dyson-series perturbation theory~\cite{seriesExpBook,FERNANDEZ2,GoldbergerAdams,RevModPhys.27.381,Taubmann}. We begin by deriving the partition function expansion.

\section{Off-diagonal expansion of the quantum partition function}
The canonical partition function of a system whose Hamiltonian is $H$ is given by 
\beq
Z=\tr \left[ \e^{-\beta H}\right]\,.
\eeq
Our decomposition begins by first writing the Hamiltonian in the form 
\beq
H = H_\D - \sum_j \Gamma_j V_j \,.
\eeq
Here, $H_\D$ is the `classical' part of the Hamiltonian, i.e., a diagonal operator in some known basis, which we refer to as the computational basis, and whose basis states are denoted by $|\bm{\sigma}\rangle$. The operators $V_j $ are off-diagonal permutation operators obeying
${V}_j | \bm{\sigma} \rangle = | \bm{\sigma}' \rangle$
for every basis state $|\bm{\sigma}\rangle$, where $ |\bm{\sigma}' \rangle$ is also a basis state. The $\Gamma_j$ will in general be diagonal operators that couple to the off-diagonal operators $V_j$. 
To avoid cluttering the derivation we shall assume that  $\Gamma_j=\Gamma \times \mathbb{1}$, where $\Gamma$ is a real-valued parameter and $\mathbb{1}$ is the identity matrix, although as will become clear shortly, the general case is no different to derive. 

We begin by replacing the trace operation with the explicit sum $\sum_{\{\bm{\sigma}\}} \langle \bm{\sigma}| \cdot | \bm{\sigma} \rangle$ and expanding the exponent in a Taylor series in the inverse temperature $\beta$, in which case the partition function can be written as 
\bea
Z &=&\sum_{\{\bm{\sigma}\}} \sum_{n=0}^{\infty}\frac{\beta^n}{n!} \langle \bm{\sigma}| (-H)^n | \bm{\sigma} \rangle \nonumber \\
&=& \sum_{\{\bm{\sigma}\}} \sum_{n=0}^{\infty}\frac{\beta^n}{n!} \langle \bm{\sigma}| (-H_\D + \Gamma \sum_j V_j)^n | \bm{\sigma} \rangle \nonumber \\
&=& \sum_{\{\bm{\sigma}\}} \sum_{n=0}^{\infty}  \sum_{\{ {S}_{n}\}} \frac{\beta^n}{n!} \langle \bm{\sigma}| {S}_{n} | \bm{\sigma} \rangle \,,
\eea
where in the last step we have also expanded $(-H)^n$, and $\{{S}_{n}\}$ denotes the set of all possible combinations of operator products $S_n$ of length $n$ consisting of products of basic operators $H_\D$ and $V_j$.

\begin{figure}[htp]
\includegraphics[width=0.99\columnwidth]{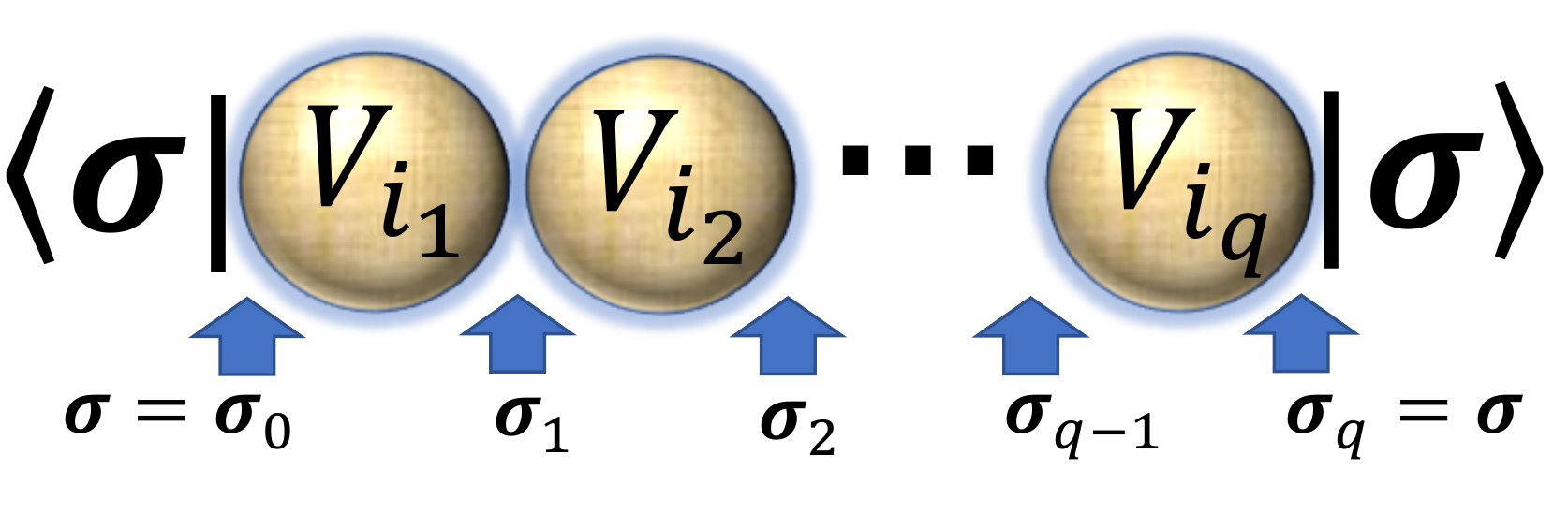}
\caption{\label{fig:sequence}{\bf A schematic representation of a $\langle \bm{\sigma}| S_q|\bm{\sigma}\rangle$ term.} The sequence of operators \hbox{${S}_{q}={V}_{i_1} \cdot {V}_{i_2} \cdots {V}_{i_q}$} is sandwiched between classical bra $\langle \bm{\sigma}|$ and ket $|\bm{\sigma}\rangle$ states, inducing a sequence of classical states ($|{\bm{\sigma}}_0\rangle, \ldots, |{\bm{\sigma}}_q\rangle$). The classical energies of the states $|{\bm{\sigma}}_i\rangle$, namely,
$E_i=E_c({\bm{\sigma}}_i)=\langle {\bm{\sigma}}_i | H_\D|{\bm{\sigma}}_i\rangle$ are the building blocks of the divided-difference weight $(-\Gamma)^q \e^{-\beta[E_0,\ldots,E_q]}$.} 
\end{figure}

As a next step, we rid the terms $\langle \bm{\sigma}| {S}_n | \bm{\sigma} \rangle$ of the diagonal Hamiltonian operators inside $S_n$ by evaluating their action on the relevant basis states, leaving only the off-diagonal permutation operators $V_j$ unevaluated inside the sequence. Lumping together all terms with the same `off-diagonal backbone,'  we arrive at 
\bea\label{eq:snsq}
Z  &=&
\sum_{\{\bm{\sigma}\}} \sum_{q=0}^{\infty} \sum_{\{ {S}_{q}\}}  \Gamma^q \langle \bm{\sigma}| {S}_q | \bm{\sigma} \rangle  \\\nonumber
&\times&\left( \sum_{n=q}^{\infty} \frac{\beta^n(-1)^{n-q}}{n!} \right. 
\left.  \sum_{\sum k_i=n-q} E^{k_0}({\bm{\sigma}}_0) \cdots E^{k_{q}}({\bm{\sigma}}_{q}) \right)\,,
\eea
where $E_c({\bm{\sigma}}_i)=\langle {\bm{\sigma}}_i |H_\D | {\bm{\sigma}}_i \rangle$ and ${\{{S}_q\}}$ denotes the set of all possible combinations of operator products $S_q=V_{i_1}\cdots V_{i_q}$ of length $q$ of \emph{off-diagonal} operators $V_j$. The expression in  parenthesis sums over the diagonal contributions of all $\langle \bm{\sigma}| {S}_n | \bm{\sigma} \rangle$ terms that correspond to a single $\langle \bm{\sigma}| {S}_q | \bm{\sigma} \rangle$ term. The various $\{{\bm{\sigma}}_j\}$ states are the states obtained from the action of the ordered $V_j$ operators in the sequence ${S}_q$ on $|{\bm{\sigma}}_0\rangle$, then on $|{\bm{\sigma}}_1\rangle$, and so forth. For ${S}_q={V}_{i_1}  \cdots {V}_{i_q}$, 
we obtain $|{\bm{\sigma}}_0\rangle=|\bm{\sigma}\rangle, {V}_{i_1}|{\bm{\sigma}}_0\rangle=|{\bm{\sigma}}_1\rangle, {V}_{i_2}|{\bm{\sigma}}_1\rangle=|{\bm{\sigma}}_2\rangle$ and so forth.
Figure~\ref{fig:sequence} provides a schematic representation of $\langle \bm{\sigma}| {S}_q | \bm{\sigma} \rangle$. 

After a change of variables, $n \to n+q$, we arrive at:
\bea
Z  &=& \sum_{\{\bm{\sigma}\}} \sum_{q=0}^{\infty} \sum_{\{ {S}_{q}\}} \langle \bm{\sigma}| {S}_q | \bm{\sigma} \rangle \times \\\nonumber 
&&\left( (\beta \Gamma)^q \sum_{n=0}^{\infty} \frac{(-\beta)^n}{(n+q)!}
\sum_{\sum k_i=n}  E^{k_0}({\bm{\sigma}}_0) \cdots E^{k_{q}}({\bm{\sigma}}_{q}) \right)\,.
\eea
Abbreviating $\x_i \equiv E_c({\bm{\sigma}}_i)$ (note that the various $\{\x_i\}$ are functions of the $|{\bm{\sigma}}_i\rangle$ states generated by the operator products ${S}_q$), the partition function becomes
\bea
Z  &=& \sum_{q=0}^{\infty}(-\Gamma)^q \sum_{z, \{ {S}_{q}\}}  \langle \bm{\sigma}| {S}_q | \bm{\sigma} \rangle 
\\\nonumber&\times&
 \left( 
\sum_{\{ k_i\}=(0,\ldots,0)}^{(\infty,\ldots,\infty)} \frac{(-\beta)^q}{(q+\sum k_i)!} \prod _{j=0}^{q} (-\beta \x_j)^{k_j} 
\right)
 \,.
 \eea
Interestingly, the infinite sum inside the parentheses can be simplified to give the \emph{exponent of divided differences} of the $\x_i$'s (we give a short description of divided differences and an accompanying proof of the above assertion in the Appendix), namely, it can be succinctly rewritten as: 
\bea
\sum_{\{ k_i\}} \frac{(-\beta)^q}{(q+\sum k_i)!} \prod _{j=0}^{q} (-\beta \x_j)^{k_j} 
=e^{-\beta[\x_0,\ldots,\x_q]} \nonumber\\\,, 
\eea
where $[\x_0,\ldots,\x_q]$ is a \emph{multiset} of energies and where a function $F[\cdot]$ of a multiset of input values is defined by
\beq 
F[\x_0,\ldots,\x_q] \equiv \sum_{j=0}^{q} \frac{F(\x_j)}{\prod_{k \neq j}(\x_j-\x_k)}
\eeq
and is called the \textit{divided differences}~\cite{dd:67,deboor:05} of the function $F[\cdot]$ with respect to the input $[\x_0,\ldots,\x_q]$. 

\begin{figure*}[htp]
\includegraphics[width=1.5\columnwidth]{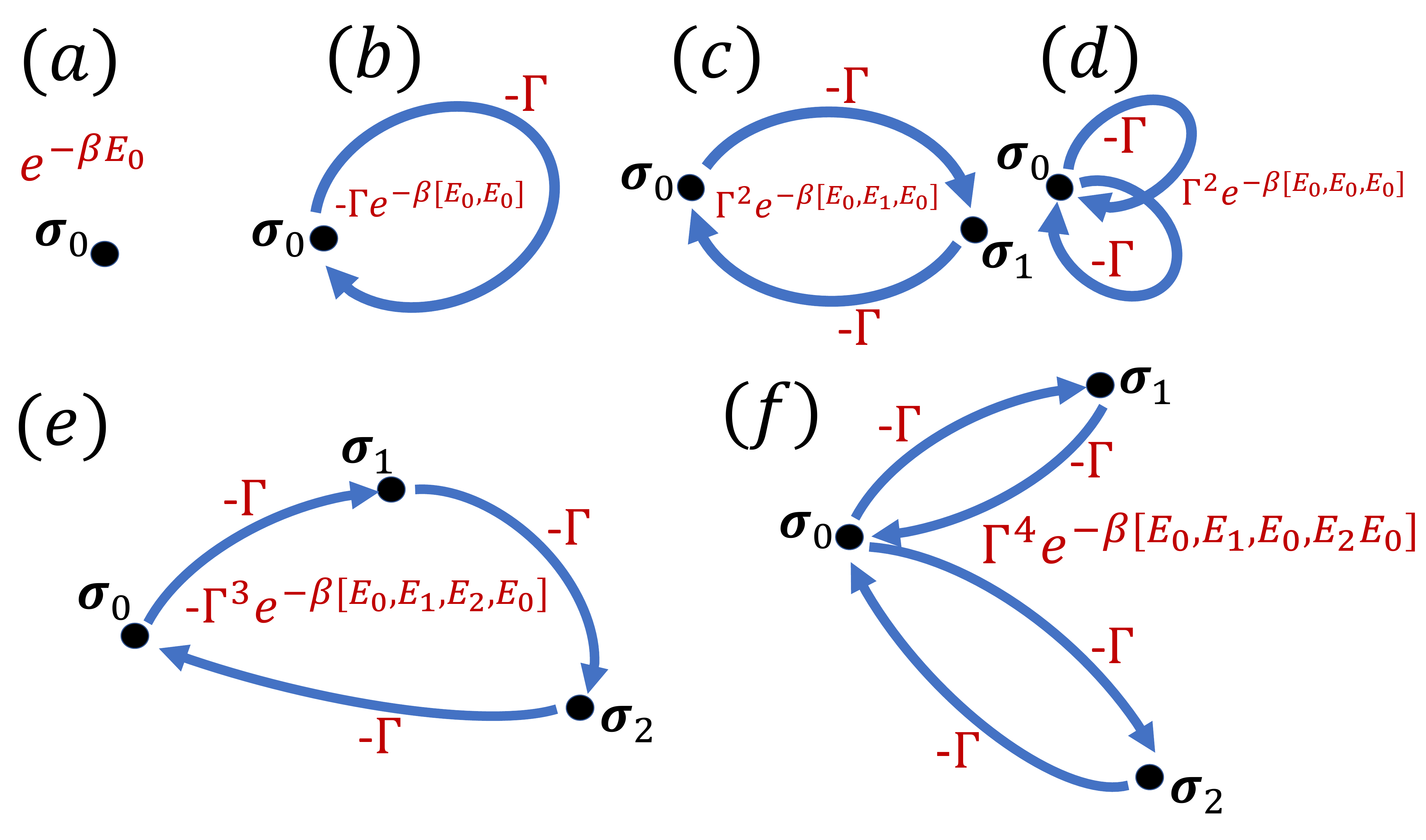}
\caption{\label{fig:qmcWeights} {\bf Diagrammatic representation of terms in the partition function expansion.} Each term is assigned a Boltzmann-weight-like divided-difference expression of the form $(-\Gamma)^q\e^{-\beta[E_0,\ldots,E_q]}$ calculated from the classical energies $E_j$ of the classical states $|\bm{\sigma}_j\rangle$. Only closed cycles contribute to the expansion. (a) A zeroth-order classical term. These terms appear in the decomposition of the classical partition function $Z_0$. Its weight is a classical Boltzmann weight. (b) First-order terms contain a single edge and two occurrences of the classical state. (c)-(d) Two examples for second-order terms. (e)-(f) Examples for third- and fourth-order terms, respectively.}
\end{figure*}

The evaluation of $F[\x_0,\ldots,\x_q]$ can conveniently be carried out using $q(q-1)/2$ operations via the recursion relations (see the Appendix for more details)
\beq\label{eq:ddr}
F[\x_i,\ldots,\x_{i+j}] 
=\frac{F[\x_{i+1},\ldots , \x_{i+j}] - F[\x_i,\ldots , \x_{i+j-1}]}{\x_{i+j}-\x_i} \,,\nonumber\\
\eeq
with $i\in\{0,\ldots,q-j\}$ and $\ j\in\{1,\ldots,q\}$,  augmented with the initial conditions
\beq\label{eq:divideddifference3}
F[\x_i] = F(\x_{i}), \qquad i \in \{ 0,\ldots,q \} \,. 
\eeq
We note that the above expression is also well-defined in cases where the inputs have repeated values, in which case one is required to take the appropriate limit in order to evaluate the function. Specifically, in the case where $\x_0=\x_1=\ldots=\x_q=\x$, the definition of divided differences reduces to: 
\beq\label{eq:derF}
F[\x_0,\ldots,\x_q] = \frac{F^{(q)}(\x)}{q!} \,,
\eeq 
where $F^{(q)}(\cdot)$ stands for the $q$th derivative of $F(\cdot)$.
The above infinite sum over energies reduces $Z$ to
\beq
Z  = \sum_{\{\bm{\sigma}\}} \sum_{q=0}^{\infty} \sum_{\{ {S}_{q}\}} \langle \bm{\sigma}| {S}_q | \bm{\sigma} \rangle (-\Gamma)^q e^{-\beta[\x_0,\ldots,\x_q]} \,.
\label{eq:SSE3}
\eeq
Furthermore, since by construction the term $\langle \bm{\sigma}| {S}_q | \bm{\sigma} \rangle$ evaluates to either $0$ or to $1$ (the operation $S_q|\bm{\sigma}\rangle$ returns a basis state $|\bm{\sigma}'\rangle$ and therefore $\langle \bm{\sigma}| S_q |\bm{\sigma}\rangle=\langle \bm{\sigma}| \bm{\sigma}'\rangle=\delta_{\bm{\sigma},\bm{\sigma}'}$), the partition function can be cast in its final form as a sum over all pairs $\{\bm{\sigma},{S}_{q}\}$ corresponding to $\langle \bm{\sigma}| {S}_q | \bm{\sigma} \rangle=1$:
\beq\label{eq:ZqZq}
Z  = \sum_q Z_q
\; \text{with} \;
Z_q = \sum_{\langle \bm{\sigma}| {S}_q | \bm{\sigma} \rangle=1}  (-\Gamma)^q e^{-\beta[\x_0,\ldots,\x_q]} \,.
\eeq
The expansion, Eq.~(\ref{eq:ZqZq}), is a series in the `quantum strength' parameter $\Gamma$ of the model. Specifically, it contains as a partial sum the classical partition function decomposition of its diagonal part $H_\D$, namely,
\beq\label{eq:Z0}
Z_0=\sum_{\{ \bm{\sigma}\}} \e^{-\beta E_c(\bm{\sigma})} \,.
\eeq
Moreover, the various summands in the partial sums $Z_q$, Eq.~(\ref{eq:ZqZq}), admit diagrammatic representations. Terms in the partial classical sum $Z_0$ generated by classical configurations $\bm{\sigma}$ and \hbox{$S_0=\mathbb{1}$} are depicted as points [shown in Fig.~\ref{fig:qmcWeights}(a)], with weights corresponding to standard Boltzmann weights. 
Quantum terms with $q>0$ correspond to loop diagrams, or cycles, originating in a classical state $\bm{\sigma}$, hopping to other classical states via the $V_j$ operators in $S_q$, each of which contributing a factor of $-\Gamma$, eventually circling back to the originating classical state, as shown in Figs.~\ref{fig:qmcWeights}(b)-(f). The number of edges in a diagram is the order of the term, $q$. 
Figure~\ref{fig:qmcWeights}(b) depicts a diagram whose order is $q=1$ for which $S_q$ contains a single $V_j$ operator. Figures~\ref{fig:qmcWeights}(c)-(d) on the other hand correspond to second-order terms containing two edges, and Figs.~\ref{fig:qmcWeights}(e)-(f) are examples of third- and fourth-order terms, respectively. The weight of each coefficient is calculated from the classical energies of the visited nodes via the divided-difference procedure. Interestingly, the order in which the nodes are visited is immaterial to the calculation of a weight.  

We next illustrate the usefulness of the derived series expansion by obtaining analytical expressions for the leading terms of the partition functions of two quantum spin models. The models we consider are the quantum 1D Ising model with longitudinal and transverse fields and the quantum 1D Heisenberg model. 

\section{The quantum 1D-Ising model\label{sec:1dIsing}}
The Hamiltonian of the quantum 1D Ising model (with assumed periodic boundary conditions) is given by 
\beq
H=-J \sum_j \sigma^z_j \sigma^z_{j+1} -h \sum_j \sigma^z_j -\Gamma \sum_{j} \sigma^x_j \,.
\eeq 
Here, the classical part of the Hamiltonian is \hbox{$H_\D=-J \sum_j \sigma^z_j \sigma^z_{j+1} -h \sum_j \sigma^z_j$} and the off-diagonal operators are $V_j=\sigma^x_j$. The computational basis states  are
$|\bm{\sigma}\rangle = \bigotimes_i |\sigma_i\rangle$ where $i=1\ldots N$, $\sigma_i=\pm 1$ denote the orientation of spin $i$ and $|\sigma_i\rangle$ denote the corresponding eigenvectors of $\sigma^z_i$. 

The zeroth-order of the partition function expansion is given by Eq.~(\ref{eq:Z0})
with the classical energy $E_c(\bm{\sigma})=\langle \bm{\sigma}| H_\D |\bm{\sigma}\rangle$:
it is simply the partition function of the classical model $H_\D$. The summands in $Z_0$ correspond to `point' contributions as depicted in Fig~\ref{fig:qmcWeights}(a). 

The classical partition function $Z_0$ can be calculated exactly by casting it in terms of the $2\times2$ transfer matrix $T$~\cite{Baxter:book} whose elements are
\bea\label{eq:T}
T_{(\sigma_1,\sigma_2)}&=& \e^{-\beta \left[-J \sigma_1 \sigma_2 -\frac{h}{2} (\sigma_1+\sigma_2)\right]} \,.
\eea
The transfer matrix can be spectrally decomposed to $T=\lambda_{-} |\phi_-\rangle\langle \phi_-|+\lambda_{+} |\phi_+\rangle\langle \phi_+|$ where
\beq
\lambda_{\pm} = \e^{\beta J} \cosh \beta h \pm  \sqrt{ \e^{-2 \beta J} + \e^{2 \beta J} \sinh \beta h}
\eeq
are its eigenvalues and 
\beq
|\phi^{\pm}\rangle=\left( \e^{2 \beta J} \sinh \beta h \pm \sqrt{1+ \e^{4 \beta J} \sinh^2 \beta h}, 1\right)
\,,
\eeq
are the corresponding (un-normalized)  eigenvectors.

In terms of the transfer matrix $T$, the classical partition function $Z_0$ evaluates to
\bea
Z_0&=&\sum_{\sigma_1 = \pm 1, \sigma_2 =\pm 1, \ldots} \prod_j \langle \sigma_j | T |\sigma_{j+1}\rangle=\tr \left( T^N\right)\nonumber\\
&=&\lambda_+^N +\lambda_{-}^N \approx \lambda_+^N \,,
\eea
where in the last step we have taken the large $N$ limit. 

The first-order term $Z_1$, which corresponds to sums of terms presented in Fig~\ref{fig:qmcWeights}(b), vanishes as it contains terms of the form $\langle \bm{\sigma}| \sigma^x_i |\bm{\sigma}\rangle$ which evaluate to zero.  Similarly, all odd-ordered coefficients vanish, explicitly, $Z_{2 k +1}=0$ for all integers $k\geq0$.

The leading quantum correction is the second-order term $Z_2$, and it contains contributions from terms depicted in Fig~\ref{fig:qmcWeights}(c)-(d). It reads
\bea
Z_2=\Gamma^2 \sum_{\langle \bm{\sigma}| \sigma^x_i \sigma^x_i | \bm{\sigma} \rangle=1}\e^{-\beta [E_c(\bm{\sigma}),E_c({\bm{\sigma}}_i),E_c(\bm{\sigma})]}
\eea
where ${\bm{\sigma}}_i$ denotes the configuration $\bm{\sigma}$ with its $i$th spin flipped. 
Evaluation of the triple-energy divided difference yields
\beq
 \e^{-\beta [E_c(\bm{\sigma}),E_c({\bm{\sigma}}_i),E_c(\bm{\sigma})]} = \e^{-\beta E_c(\bm{\sigma})} g(\Delta E_i) \,,
 \eeq
 where 
 \bea
 \Delta E_i= E_c(\bm{\sigma})-E_c({\bm{\sigma}}_i) 
= 2 \sigma_i(J \sigma_{i-1}+ J \sigma_{i+1}+h) \,,\nonumber\\
 \eea
 and 
 \beq \label{gde}
 g(\Delta E)= \frac{\beta} {\Delta E} -\frac{1} {(\Delta E)^2}+\frac{\e^{-\beta \Delta E}} {(\Delta E)^2}\,.
 \eeq
Importantly, $g(\Delta E)$ is also well-defined in the limit of small $\Delta E$, namely, $\lim_{\Delta E \to 0}  g(\Delta E)=\beta^2/2$ [see Eq.~(\ref{eq:derF})]. 

Taking advantage of the translational symmetry of the model, we evaluate $Z_2$ by calculating the contribution from a single $\sigma^x_i \sigma^x_i$ pair for an arbitrary spin index $i$ and multiply the end result by the number of spins $N$. To that aim, we split the sum to eight different partial sums, corresponding to the number of  combined orientations of the triplet of spins $\sigma_{i-1}, \sigma_{i}$ and $\sigma_{i+1}$.
We thus write the second-order term $Z_2$ as
\beq
Z_2=  N \sum_{\substack{
\sigma_{i-1} = \pm 1 \\ \phantom{Al}\sigma_{i} =\pm 1 \\ \sigma_{i+1} =\pm 1}
} Z^{\sigma_{(i-1, i, i+1)}}_2
\eeq 
where for any configuration $\bm{\sigma}$ with fixed $\sigma_{i-1},\sigma_{i},\sigma_{i+1}$, we have
\beq
Z^{\sigma_{(i-1, i, i+1)}}_2=\Gamma^2 g(\Delta E_{i}) \sum_{\bm{\sigma}\vert \sigma_{(i-1, i, i+1)}} \e^{-\beta E_c(\bm{\sigma})} 
\eeq
where the sum over $\bm{\sigma}\vert \sigma_{(i-1, i, i+1)}$ denotes summation over all configurations with the spins $\sigma_{i-1}, \sigma_{i}$ and $\sigma_{i+1}$ fixed. 
\begin{widetext}
Similar to $Z_0$, the above sum may be cast in terms of transfer matrix elements:
\bea
Z^{\sigma_{(i-1, i, i+1)}}_2= \Gamma^2 g(\Delta E_i)  
T_{(\sigma_{i-1},\sigma_{i})} T_{(\sigma_{i},\sigma_{i+1})}\left( T^{N-2}\right)_{(\sigma_{i+1},\sigma_{i-1})} \,.
\eea
Summing over all eight possible orientations of the three spins, the second-order coefficient becomes in the large $N$ limit
\bea
Z_2&=&  Z_0  \times \left[ N \Gamma^2\lambda^{-2}_{+} \sum_{\substack{
\sigma_{i-1} = \pm 1 \\ \phantom{Al}\sigma_{i} =\pm 1 \\ \sigma_{i+1} =\pm 1}
} g(\Delta E_i)   
 \e^{-\beta \left[-J \sigma_{i}( \sigma_{i-1}+ \sigma_{i+1}) -\frac{h}{2} (\sigma_{i-1}+2\sigma_i +\sigma_{i+1}) \right]}  \phi^{+}_{(\sigma_{i+1})}\phi^{+}_{(\sigma_{i-1})} \right]
\,,
\eea
where $\phi^{+}_{(\sigma_{i \pm 1})} $ is the $(\sigma_{i \pm 1})$th element of $|\phi^+\rangle$. 
\end{widetext}

Having calculated $Z_2$, we obtained an analytical expression for the quantum partition function of the 1D Ising model with longitudinal and transverse fields to third order in $\Gamma$.
From this expression, one may easily calculate, to that order, thermal averages for various physical quantities at arbitrary values of inverse-temperature $\beta$. 

The quantum 1D Ising model can alternatively be expanded in $J$ and $h$ rather than in $\Gamma$ if the $z$ and $x$ bases are swapped, i.e., if $H$ is written as: 
\beq
H=-J \sum_j \sigma^x_j \sigma^x_{j+1} -h \sum_j \sigma^x_j -\Gamma \sum_{j} \sigma^z_j \,.
\eeq 
Here the classical energy is \hbox{$H_\D=-\Gamma \sum_{j} \sigma^z_j$} and there are two types of off-diagonal operators 
\hbox{$V_j^{(h)}=\sigma^x_j$} and \hbox{$V_j^{(J)}=\sigma^x_j \sigma^x_{j+1}$}. 

In this case, the classical partition function $Z_0$ decouples to a product of $N$ single-spin functions
\beq
Z_0=\left( \e^{-\beta \Gamma} +  \e^{\beta \Gamma} \right)^N = 2^N \cosh^N \beta \Gamma\,,
\eeq
and the first-order terms \hbox{$\langle  \bm{\sigma} | V_j^{(h)}|\bm{\sigma}\rangle$} and \hbox{$\langle  \bm{\sigma} | V_j^{(J)}|\bm{\sigma}\rangle$} vanish as before to give $Z_1=0$. 

To calculate $Z_2$, one must consider two types of contributions, with diagrammatic representations as depicted in Fig~\ref{fig:qmcWeights}(c)-(d) for the two types of off-diagonal terms. 
The first is
\bea
Z^{(h)}_2&=&h^2 \sum_i \langle \bm{\sigma}| \sigma^x_i \sigma^x_i |\bm{\sigma}\rangle \e^{-\beta[E_c(\bm{\sigma}),E_c({\bm{\sigma}}_i),E_c(\bm{\sigma})]} \,,\nonumber\\
\eea
where as before ${\bm{\sigma}}_i$ denotes the configuration ${\bm{\sigma}}$ with its $i$th spin flipped.
For any given spin index $i$, the triple-energy divided-difference weight evaluates to:
\beq
\e^{-\beta[E_c(\bm{\sigma}),E_c({\bm{\sigma}}_j),E_c(\bm{\sigma})]}= \e^{-\beta E_c(\bm{\sigma})} g(\Delta E_i)
\eeq
where $\Delta E_i=2 \Gamma \sigma_i$. Simplifying the expression, we obtain
\beq
Z^{(h)}_2 = \sum_i \sum_{\sigma_i=\pm 1} g(\Delta E_i)\sum_{\sigma_{j\neq i}=\pm 1} \e^{-\beta E_c(\bm{\sigma})} \,.
\eeq
Employing the transfer matrix trick again, we arrive at the explicit expression
\bea
Z_2^{(h)}&=&\frac{h^2 \beta}{\Gamma} N 2^{N-1} \cosh^{N-1}\beta \Gamma \sinh \beta \Gamma \nonumber\\
&=&Z_0 \times \frac{N h^2 \beta}{2 \Gamma} \tanh \beta \Gamma \,.
\eea
The second contribution to $Z_2$ comes from pairs of $V_j^{(J)}$ operators. 
Denoting by $E_c({\bm{\sigma}}_j)$ the classical energy of the configuration $\bm{\sigma}$ with both its $j$th and $(j+1)$th spins flipped, i.e., $E_c({\bm{\sigma}}_j)=E_c({\bm{\sigma}})+2 \Gamma(\sigma_j + \sigma_{j+1})$,
the $Z_2^{(J)}$ term evaluates to
\bea
Z_2^{(J)}&=&J^2 \sum_j \langle \bm{\sigma}| \sigma^x_j \sigma^x_{j+1}  \sigma^x_j \sigma^x_{j+1} |\bm{\sigma}\rangle \e^{-\beta[E_c(\bm{\sigma}),E_c({\bm{\sigma}}_j),E_c(\bm{\sigma})]}
\nonumber\\
&=& \frac{J^2 \beta}{\Gamma} N 2^{N-3} \cosh^{N-2} \beta \Gamma \left(2 \beta \Gamma + \sinh 2\beta \Gamma \right)
\nonumber\\
&=& Z_0 \times \frac{N J^2 \beta}{8 \Gamma} \frac{ 2 \beta \Gamma + \sinh 2\beta \Gamma}{\cosh^2 \beta \Gamma}
\,.
\eea

The third-order term $Z_3$ consists of diagrams of the form sketched in Fig.~\ref{fig:qmcWeights}(e). Here, the only non-vanishing terms consist of a single $V_j^{(J)}$ operator and two $V_j^{(h)}$ operators and sum up to 
\bea
Z_3&=&-6 h^2 J  \sum_j \langle \bm{\sigma}| \sigma^x_j \sigma^x_{j+1}  \sigma^x_j \sigma^x_{j+1} |\bm{\sigma}\rangle \e^{-\beta[E_c(\bm{\sigma}),E_c({\bm{\sigma}}_j),E_c(\bm{\sigma})]}\nonumber \\
&=& \frac{6 h^2 J}{\Gamma^3} N 2^{N-5} \cosh^{N-2} \beta \Gamma  \left[ 2 \beta \Gamma\left( \cosh 2 \beta \Gamma -2 \right) +\sinh 2 \beta \Gamma \right]\nonumber \\
&=& Z_0 \times \frac{3N h^2 J}{16 \Gamma^3}\frac{2 \beta \Gamma \left( \cosh 2 \beta \Gamma -2 \right) +\sinh 2 \beta \Gamma}{ \cosh^{2} \beta \Gamma }\,,
\eea
where an extra factor of 6 comes from the $3!$ ways in which the three operators can be arranged to form a non-vanishing $\sigma^x_j \sigma^x_{j+1}  \sigma^x_j \sigma^x_{j+1}$ sequence. 

\section{The quantum 1D-Heisenberg model\label{sec:Heisenberg}}
The next model we consider is the quantum 1D Heisenberg model whose Hamiltonian is given by
\beq
H=-\frac{\Gamma}{2} \sum_i \left( \sigma^x_i \sigma^x_{i+1} + \sigma^y_i \sigma^y_{i+1} + \sigma^{z}_i \sigma^z_{i+1}\right) 
+h  \sum_{i} \sigma^{z}_i \,,
\eeq 
where, as before, periodic boundary conditions are assumed.
Denoting for convenience the identity matrix acting on spin $i$ by $\mathbb{1}_i$, we first rewrite the Hamiltonian as
\bea\label{eq:Heisen2}
H&=&-\frac{\Gamma}{2} \sum_{i} \left( \sigma^x_i \sigma^x_{i+1} + \sigma^y_i \sigma^y_{i+1} + \sigma^{z}_i \sigma^z_{i+1} + \mathbb{1}_i \mathbb{1}_{i+1} \right)  \nonumber\\
&+& h \sum_{i} \sigma^{z}_i +\frac{\Gamma N}{2} \,.
\eea
We identify the first line as a sum of permutation operators 
\beq
V_i = \frac1{2} \sum_i \left( \sigma^x_i \sigma^x_{i+1}+ \sigma^y_i \sigma^y_{i+1} + \sigma^{z}_i \sigma^z_{i+1}+ \mathbb{1}_i \mathbb{1}_{i+1} \right) \,,
\eeq
and note that $V_i$ swaps the orientations of the $i$th and $(i+1)$th spins. 

The quantum 1D-Heisenberg model conserves magnetization in the $z$-direction, where the operator associated with the symmetry is $M_z = \sum_i \sigma^z_i$. We will focus on the zero-magnetization sector $M_z=0$ in which case the classical energy of every configuration [the second line in Eq.~(\ref{eq:Heisen2})] is constant, explicitly, $E_c(\bm{\sigma})=N \Gamma/2$. 

From Eq.~(\ref{eq:derF}) we find that constant-energy divided-differences may be recast as
\beq
(-\Gamma)^q \e^{-\beta [E_0, E_0,\ldots, E_0]}= \frac{(\beta \Gamma)^q}{q!} \e^{-\beta E_0}\,.
\eeq
It follows then that the partition function can be written as 
\beq
Z=\sum_q Z_q = \sum_q  \frac{(\beta \Gamma)^q}{q!} \e^{-\beta N \Gamma/2} \times N_q \,,
\eeq
where
\beq
N_q=\sum_{\{ \bm{\sigma} : M_z=0\}} \sum_{\{S_q\}}  \langle \bm{\sigma}| S_q | \bm{\sigma}\rangle
\eeq
counts the number of all non-vanishing $\langle \bm{\sigma}| S_q | \bm{\sigma}\rangle $ terms.

Starting with the zeroth-order contributions [Fig~\ref{fig:qmcWeights}(a)], we obtain trivially:
\beq
N_0 = \sum_{\{ \bm{\sigma} : M_z=0\}} \langle \bm{\sigma}| \bm{\sigma}\rangle = {N \choose N/2} \,,
\eeq
the number of configurations with equal number of spins pointing up and down.

The sums in the first-order term $Z_1$, corresponding to terms as in Fig~\ref{fig:qmcWeights}(b),  evaluate to
\beq
N_1=\sum_{\{ \bm{\sigma} : M_z=0\}} \sum_i \langle \bm{\sigma}| V_i | \bm{\sigma}\rangle = 2 {N-2 \choose N/2-2} \times N \,,
\eeq
enumerating all the configurations for which a swap operator leaves the configuration unchanged, i.e., $2 {N-2 \choose N/2-2}$ times the number of swap operators $N$. 

For $q=2$, the terms in $Z_2$ correspond to diagrams as depicted in Fig~\ref{fig:qmcWeights}(c)-(d) and for which
\beq
N_2=\sum_{\{ \bm{\sigma} : M_z=0\}} \sum_i \sum_j \langle \bm{\sigma}| V_i V_j | \bm{\sigma}\rangle \,.
\eeq
Here, we distinguish between three cases. In the first $j=i$, in which case, 
\beq
\sum_{\{ \bm{\sigma} : M_z=0\}} \sum_i \langle \bm{\sigma}| V_i V_i| \bm{\sigma}\rangle= {N \choose N/2} \times N \,.
\eeq
since $V_i^2=\mathbb{1}$ and the factor $N$ is the number of swap operators. The second case is one where $j=i\pm1$. Here, 
\beq
\sum_{\{ \bm{\sigma} : M_z=0\}} \sum_{i,j=i\pm1} \langle \bm{\sigma}| V_i V_j | \bm{\sigma}\rangle = 2 {N-3 \choose N/2-3} \times 2N \,.
\eeq
The terms $\langle \bm{\sigma}| V_i V_{j=i+1} | \bm{\sigma}\rangle$ and $\langle \bm{\sigma}| V_i V_{j=i-1} | \bm{\sigma}\rangle$ do not vanish if and only if the three swapped spins, $\{i, i+1,i+2\}$ and $\{i-1, i,i+1\}$ respectively, are aligned.
The number of configurations that satisfy the above condition is $2 {N-3 \choose N/2-3}$.

The third case is one in which $i$ and $j$ are non-identical and non-consecutive in which case the two operators $V_i$ and $V_j$ commute. Here, the $i$th and $(i+1)$th spins must be aligned and similarly the $j$th and $(j+1)$th spins, yielding
\bea
&&\sum_{\{ \bm{\sigma} : M_z=0\}} \sum_i \sum_{j \neq i,i\pm1} \langle \bm{\sigma}| V_i V_j | \bm{\sigma}\rangle \\\nonumber
&=&2 \left[ {N-4 \choose N/2-2}+{N-4 \choose N/2-4}\right] \times N(N-3) \,.
\eea

The $Z_3$ term is proportional to 
\beq\label{eq:Z3}
N_3=\sum_{\{ \bm{\sigma} : M_z=0\}} \sum_i \sum_j \sum_k \langle \bm{\sigma}| V_i V_j V_k | \bm{\sigma}\rangle \,.
\eeq
Here, there are four different cases to consider: i) Two (or all three) of the indices are identical; 
ii) All three indices $i, j$ and $k$ are consecutive; iii) Only two of the indices are consecutive; and iv) all three indices are at least one spin apart, in which case $V_i, V_j$ and $V_k$ all commute. 

Table~\ref{tab1} summarizes the enumeration all of non-vanishing $\langle \bm{\sigma}| S_q | \bm{\sigma}\rangle$ terms, to third order, listing the number of configurations $\bm{\sigma}$ leading to $\langle \bm{\sigma}| S_q | \bm{\sigma}\rangle=1$ terms for the various relative orderings of the swap operators.
\begin{table*}[htp]
\begin{tabular}{|c|c||c|c|}
\hline
Expansion & Relative ordering     & Number of                & Number of non-vanishing \\
 order & of the indices & operator sequences & terms per ordering\\
 \hline
  $N_0$ & --- & 1 & ${N \choose N/2}$ \\
  \hline
  $N_1$ & --- & $N$ & $2 {N-2 \choose N/2-2} $ \\
  \hline 
 &The two indices are identical & $N$ & ${N \choose N/2}$ \\
$N_2$ &The two indices are consecutive & $2N$ & $2 {N-3 \choose N/2-3}$ \\
&Indices are neither identical nor consecutive & $N(N-3)$ & $2 \left[ {N-4 \choose N/2-2}+{N-4 \choose N/2-4}\right]$ \\
  \hline
&Two or all three of the indices are identical & $N(3 N-2)$ & $2 {N-2 \choose N/2-2}$ \\
$N_3$ &The three indices are consecutive & $6 N$ & $2{N-4 \choose N/2-4}$ \\
&Only two of the indices are consecutive & $6 N(N-4)$ & $2\left[ {N-5 \choose N/2-3} +  {N-5 \choose N/2-5} \right]$ \\
&The three indices are nonidentical and nonconsecutive & $N(N-4)(N-5)$ & $2\left[ 3 {N-6 \choose N/2-4} +  {N-6 \choose N/2-6} \right]$ \\
\hline
\end{tabular}
\caption{\label{tab1}{\bf Enumeration of non-vanishing terms $\langle \bm{\sigma}| S_q | \bm{\sigma}\rangle$ for the $N$-spin quantum 1D Heisenberg model in the zero-magnetization sector.} Depending on the relative ordering of the indices, a different number of configurations $\bm{\sigma}$ will lead to non-vanishing contributions $\langle \bm{\sigma}| S_q | \bm{\sigma}\rangle=1$.}
\end{table*}
Grouping together all of the expressions computed above, we obtain an analytical expression, to third order in $\Gamma$, for the partition function of the quantum 1D Heisenberg model:
\begin{widetext}
\bea
Z= {N \choose N/2}  \e^{-\beta N \Gamma/2} 
\left[ 1 +(\beta \Gamma) \frac{N(N-2)}{2(N-1)} + \frac{(\beta \Gamma)^2}{2} \frac{N^3}{4(N-1)} + \frac{(\beta \Gamma)^3}{6}  \frac{N^2 (N-2)(N^2+4N -22)}{8 (N-1)(N-3)} +{\cal O} (\Gamma^4) \right] \,.\nonumber\\
\eea
\end{widetext}
Higher-order terms can just as easily be computed. 

\section{Relation to Thermodynamic Perturbation Theory\label{sec:schwinger}}
For completeness, in what follows we show how the series expansion proposed here is also derivable from a Dyson-series expansion of the partition function~\cite{dyson}.
We expand the partition function of $H=H_\D-\Gamma \sum_j V_j$
in the expansion parameter $\Gamma$. We begin by observing that~\cite{seriesExpBook}
\bea
&&\e^{-\beta (H_\D-\Gamma \sum_j V_j)} =e^{-\beta H_\D}\sum_{q=0}^{\infty}\Gamma^q \times \\\nonumber
&& \int_0^{\beta} \rmd \tau_1 \int_0^{\tau_1} \rmd \tau_2 \cdots \int_0^{\tau_{q-1}} \rmd \tau_q \prod_{j=1}^q \overset{\sim}{V}(\tau_j) \,,
\eea
where we have defined 
\beq
\overset{\sim}{V}(\tau_j) =\e^{\tau_j H_\D} \left( \sum_j V_j\right) \e^{-\tau_j H_\D} \,.
\eeq
Taking the trace $\sum \langle \bm{\sigma}| \cdot |\bm{\sigma}\rangle$, the $q$th-order term reads:
\bea
Z_q &=& \Gamma^q \sum_{\{ \bm{\sigma}, S_q \}} \int_0^{\beta} \rmd \tau_1 \int_0^{\tau_1} \rmd \tau_2 \cdots \int_0^{\tau_{q-1}} \rmd \tau_q \\\nonumber
&\times& \langle  \bm{\sigma}| \e^{-(\beta-\tau_1) H_\D} V_{i_1} \e^{-(\tau_1-\tau_2) H_\D} \cdots V_{i_q} \e^{-\tau_q  H_\D}  |\bm{\sigma}\rangle \,.
\eea
Acting with the exponentials on the classical $|\bm{\sigma}\rangle$ states generated by the off-diagonal permutation operators, we obtain
\bea\label{eq:zq3}
Z_q &=& \Gamma^q\sum_{\{ \bm{\sigma}\},\{S_q \}} \langle \bm{\sigma}|  V_{i_1} \cdots V_{i_q} |\bm{\sigma}\rangle \int_0^{\beta}\rmd \tau_1  \cdots \int_0^{\tau_{q-1}} \rmd \tau_q \nonumber\\
&\times&\e^{ -(\beta-\tau_1) E_0-(\tau_1-\tau_2) E_1 \cdots -(\tau_q-\tau_{q-1}) E_q} \,,
\eea
where ${\{{S}_q\}}$ denotes as before the set of all possible combinations of operator products $S_q=V_{i_1}\cdots V_{i_q}$ of length $q$ of off-diagonal operators. 
A simple change of variables  $\tau_i \to t_i/\beta$ yields
\bea\label{eq:zq4}
Z_q &=&(\beta \Gamma)^q \sum_{\{ \bm{\sigma}, S_q \}} \langle \bm{\sigma}|  V_{i_1} \cdots V_{i_q} |\bm{\sigma}\rangle \times \\\nonumber
&&  \int_0^{\beta}\rmd t_1  \cdots \int_0^{\tau_{q-1}} \rmd t_q\e^{ -\beta \left[E_0 (1-\sum_i t_i)+ E_1 t1 \cdots +E_q t_q\right]} \,.
\eea
To carry out the integration, we invoke the Hermite-Genocchi formula for functions of divided differences~\cite{deboor:05} which reads for an arbitrary function $F(\cdot)$:
\beq\label{eq:hg}
F[\x_0,\ldots,\x_n]=\int_{\Omega_n}F^{(n)}(\x_0 t_0 + \ldots \x_q t_q) \rmd t_1 \cdots \rmd t_q
\eeq
where $F^{(n)}(\cdot)$ denotes the $n$th derivative of $F(\cdot)$ and $t_0=1-\sum_i t_i$, and the volume of integration is 
\beq
\Omega_n =\left\{ (t_1,\ldots,t_n): t_i>0 \quad\text{and} \quad1-\sum_i t_i>0\right\} \,.
\eeq
Taking $F[\cdot ]$ to be $\e^{-\beta [\cdot ]}$, the right-hand side of Eq.(\ref{eq:hg}) simplifies neatly to the integral of Eq.(\ref{eq:zq4}) yielding \beq
Z_q = (-\Gamma)^q \sum_{\langle \bm{\sigma}| {S}_q | \bm{\sigma} \rangle=1} \e^{-\beta [E_0,\ldots,E_q]}
\eeq
as desired.
\section{Summary and discussion\label{sec:conclusions}}
We introduced an integral-free thermodynamic perturbation series expansion for quantum partition functions. The expansion is carried out around the partition function of the classical component of the Hamiltonian with the expansion parameter being the strength of the off-diagonal, or quantum, portion of the Hamiltonian. The proposed scheme allows for an analytical term-by-term calculation of the coefficients of the expansion, which admit simple diagrammatic representations. 

The expansion presented here has several attractive features. The calculated coefficients are complete functions of $\beta$, hence the series can be successfully used at arbitrary temperatures, yielding information about all temperature-separated phases of the system being studied. This is in contrast to phase transitions in the expansion parameter which may, on the other hand, have limited radius of convergence. The closer the system is to being classical, the more accurate the approximation is. 

As was demonstrated, differently from standard thermodynamic perturbation theory~\cite{seriesExpBook,LL:SP1}, the method derived here is neither defined with nor requires (the sometimes cumbersome) multidimensional integration of operators in imaginary time. 
We therefore hope that the scheme presented here may prove to be a useful tool in the study of quantum many-body systems. In this regard, it would be interesting to explore other quantum models that do not admit closed-form solutions, but on the other hand do allow for a useful analytical term-by-term calculation as a series, similarly to the examples worked out above.

Another use for the expansion, which has not been explored here, is towards the \emph{numerical} estimation of the different partition-function coefficients $Z_q$ using importance sampling---i.e., Monte Carlo techniques. An efficient numerical evaluation of the various terms may be useful, especially at large orders of the expansion for which the analytical treatment becomes tedious. In that context, it is interesting to observe that irrespective of the physical model being studied, the sums comprising the $Z_q$ coefficients do not suffer from the infamous sign problem~\cite{Loh-PRB-90,Troyer2005} as they are sums of the divided-difference weights $(-\Gamma)^q e^{-\beta[\x_0,\ldots,\x_q]}$, which are either strictly positive or strictly negative depending on the sign of $\Gamma$ and the parity of $q$.

\begin{acknowledgments}
We thank Lorenzo Campos-Venuti, Victor Martin-Mayor and Paolo Zanardi for useful discussions. 
The research is based upon work (partially) supported by the Office of
the Director of National Intelligence (ODNI), Intelligence Advanced
Research Projects Activity (IARPA), via the U.S. Army Research Office
contract W911NF-17-C-0050. The views and conclusions contained herein are
those of the authors and should not be interpreted as necessarily
representing the official policies or endorsements, either expressed or
implied, of the ODNI, IARPA, or the U.S. Government. The U.S. Government
is authorized to reproduce and distribute reprints for Governmental
purposes notwithstanding any copyright annotation thereon.
\end{acknowledgments}

\bibliography{refs}
\appendix

\section{Divided differences\label{sec:DividedDifference}}

We provide below a brief summary of the concept of divided differences which is a recursive division process. This method is typically encountered when calculating the coefficients in the interpolation polynomial in the Newton form.

The divided differences~\cite{dd:67,deboor:05} of a function $F(\cdot)$ is defined as
\beq\label{eq:divideddifference2}
F[x_0,\ldots,x_q] \equiv \sum_{j=0}^{q} \frac{F(x_j)}{\prod_{k \neq j}(x_j-x_k)}
\eeq
with respect to its input values $[x_0,\ldots,x_q]$. The above expression is well-defined even if the inputs have repeated values, in which case one must resort to a limiting process. Specifically, in the case where $x_0=x_1=\ldots=x_q=x$, the definition of divided differences reduces to: 
\beq\label{eq:derF2}
F[x_0,\ldots,x_q] = \frac{F^{(q)}(x)}{q!} \,,
\eeq 
where $F^{(q)}(\cdot)$ stands for the $q$th derivative of $F(\cdot)$.

A divided difference can alternatively be defined via the following recursion relations which also provide.a simple way to evaluate it. 
\bea\label{eq:ddr2}
&&F[x_i,\ldots,x_{i+j}] \\\nonumber
&=& \frac{F[x_{i+1},\ldots , x_{i+j}] - F[x_i,\ldots , x_{i+j-1}]}{x_{i+j}-x_i} \,,
\eea 
with $i\in\{0,\ldots,q-j\},\ j\in\{1,\ldots,q\}$ with the initial conditions
\beq\label{eq:divideddifference4}
F[x_i] = F(x_{i}), \qquad i \in \{ 0,\ldots,q \}  \quad \forall i \,.
\eeq
A function of divided differences can be defined in terms of its Taylor expansion. In the case where $F(x)=\e^{-\beta x}$, we have
\beq
\e^{-\beta [x_0,\ldots,x_q]} = \sum_{n=0}^{\infty} \frac{(-\beta)^n [x_0,\ldots,x_q]^n}{n!} \ . 
\eeq 
Moreover, it is easy to verify that
\beq \nonumber 
[x_0,\ldots,x_q]^{q+m} = \Bigg\{ 
\begin{tabular}{ l c l }
  $m<0$ & \phantom{$0$} & $0$ \\
  $m=0$ & \phantom{$0$} & $1$ \\
  $m>0$ & \phantom{$0$} & $\sum_{\sum k_j = m} \prod _{j=0}^{q} x_j^{k_j}$ \\
\end{tabular}
 \,.
\eeq
One may therefore write:
\bea
\e^{-\beta[x_0,\ldots,x_q]} &=& \sum_{n=0}^{\infty} \frac{(-\beta)^n [x_0,\ldots,x_q]^n}{n!}\\
&=&\sum_{n=q}^{\infty} \frac{(-\beta)^n [x_0,\ldots,x_q]^n}{n!} \nonumber\\
&=& 
\sum_{m=0}^{\infty} \frac{(-\beta)^{q+m} [x_0,\ldots,x_q]^{q+m}}{(q+m)!}\nonumber\\
&=&\sum_{m=0}^{\infty} \frac{(-\beta)^q}{(q+m)!} \sum_{\sum k_j = m} \prod _{j=0}^{q} (-\beta x_j)^{k_j}\nonumber
 \,.
\eea
as was asserted in the main text.
\end{document}